\documentclass[12pt, centerh1]{article} 

\usepackage{amssymb,natbib}
\usepackage{amsfonts, amsmath, amssymb, marvosym,colonequals,algorithm}
\usepackage{algorithm,xcolor}
\usepackage{bm,relsize}
\usepackage{geometry}
\usepackage{lscape}
\usepackage{mathtools}
\usepackage{caption}
\usepackage{theorem}
\usepackage{placeins}
\usepackage{hyperref}
\usepackage{enumerate}

\newcommand{\ident}{\mathbf{I}}
\newcommand{\vecx}{\mathbf{x}}

\newcommand{\vecz}{\mathbf{z}}

\newcommand{\vecpi}{\mbox{\boldmath$\pi$}}

\newcommand{\vecmu}{\mbox{\boldmath$\mu$}}

\newcommand{\vectheta}{\mbox{\boldmath$\theta$}}
\newcommand{\varthet}{\mbox{\boldmath$\vartheta$}}

\newcommand{\matsig}{\mathbf{\Sigma}}

\title{An Evolutionary Algorithm with Crossover and Mutation for Model-Based Clustering}
  \author{Sharon M.\ McNicholas$^{*}$, Paul D.\ McNicholas$^{*}$ and Daniel A. Ashlock$^{**}$}
  \date{\small $^{*}$Department of Mathematics and Statistics, McMaster University, Ontario, Canada.\\
 $^{**}$Department of Mathematics and Statistics, University of Guelph, Ontario, Canada.}

\begin{document}

\maketitle

\begin{abstract}
An evolutionary algorithm (EA) is developed as an alternative to the EM algorithm for parameter estimation in model-based clustering. This EA facilitates a different search of the fitness landscape, i.e., the likelihood surface, utilizing both crossover and mutation. Furthermore, this EA represents an efficient approach to ``hard" model-based clustering and so it can be viewed as a sort of generalization of the $k$-means algorithm, which is itself equivalent to a restricted Gaussian mixture model. The EA is illustrated on several datasets, and its performance is compared to other hard clustering approaches and model-based clustering via the EM algorithm.\\

\noindent \textbf{Keywords}: clustering; crossover; evolutionary algorithm; mixture models; mutation; model-based clustering.
\end{abstract}

\section{Introduction}\label{sec:intro}

Model-based clustering is the use of mixture models for clustering. Recent reviews of model-based clustering are given by \cite{bouveyron14} and \cite{mcnicholas16b}, while extensive details on mixture models are provided by \cite{titterington85} and \cite{mclachlan00b}. The expectation-maximization (EM) algorithm \citep{dempster77} is commonly used to estimate parameters in model-based clustering problems; however, it is prone to becoming stuck at local maxima \citep[see, e.g.,][]{titterington85}. As an alternative to using the EM algorithm to estimate parameters, an evolutionary algorithm (EA) is developed herein. This facilitates a different search of the fitness landscape (i.e., the likelihood surface) and can be viewed as an algorithm for ``hard" model-based clustering. 
Let \begin{equation}\label{eqn:thisone1}z_{ig} = \begin{cases}1 & \text{if observation}~\vecx_i \text{ belongs to component}~g,\\
0 & \text{otherwise},
\end{cases}\end{equation} so that the goal of model-based clustering is to predict $z_{ig}$ for each ($p$-dimensional) observation $\vecx_i$ and each component~$g$. In this context, ``hard'' means that the prediction for $z_{ig}$ is restricted to taking a value $0$ or $1$ and, for convenience, we use the notation $\tilde{z}_{ig}\in\{0,1\}$ to denote the prediction for $z_{ig}$ arising from our EA. This differs from the typical EM approach, where the prediction is a probability $\hat{z}_{ig}\in[0,1]$ which may be reported as is (soft) or subjected to an \textit{a~posteriori} hardening. Our EA can also be viewed as a sort of generalization of the $k$-means algorithm. \cite{celeux92} show that the $k$-means algorithm is equivalent to a classification EM (CEM) algorithm for a Gaussian mixture model with equal mixing proportions and common spherical component covariances (i.e., component covariance matrices $\matsig_g=\lambda\ident_p$, where $\lambda\in\mathbb{R}^+$ and $\ident_p$ is the $p$-dimensional identity matrix) --- \cite{vermunt11} also presents an argument in this direction. Some historical context and further details on the CEM algorithm are provided by \cite{mclachlan82}.

\section{Background}
\subsection{Evolutionary Algorithms}

Evolutionary computation is a paradigm in which a computer algorithm incorporates some of the elements of the biological theory of evolution. Evolutionary operations are performed on members of a population, which reproduce and create new population members. The new members replace less ``fit" members from the previous generation, and the process is continued until some stopping criterion is met. Evolutionary operations include actions such as crossover and mutation. The measure of ``fitness" used is determined by the goal of the evolution, i.e., the function that is being optimized. 
The field of evolutionary computation is interdisciplinary, with practitioners approaching it from a variety of different perspectives such as computer science, biology, and statistics. This leads to a situation where terminology is often ambiguous; the following explanation of the basic terminology used herein is based on the concepts as outlined in \cite{Ashlock1}.

Evolution occurs when a population is subject to change over time. Population members from each generation undergo a selection process.
Crossover is the combining of two data structures to produce at least one new structure. 
Two-point crossover occurs when two data structures are selected to be parents. Then, two positions (the same for each structure) are selected at random and the genetic material between these two positions is exchanged between the parents (see Table~\ref{2pcimage}). The resulting structures are called offspring, or children.
Mutation is the process by which random changes are made to a population member's structure and can be used to produce a constant supply of minor variation in the population over time. In fitness-based reproduction, solutions that are deemed to be fitter, based on a predetermined fitness function, are preferentially selected to reproduce.
\begin{table}[!ht]
\caption{Illustration of 2-point crossover, similar to that used by \cite{Ashlock1}.}
\label{2pcimage}
\vspace{-0.12in}
\centering\begin{tabular*}{0.9\textwidth}{@{\extracolsep{\fill}}lc}\hline
Parent~1 &xxxxxxxxxxxxxxxxxxxxxx\\
Parent~2& yyyyyyyyyyyyyyyyyyyyyy\\\hline
Child~1& xxxxxxxxyyyyyyyxxxxxxx\\
Child~2& yyyyyyyyxxxxxxxyyyyyyy\\\hline
\end{tabular*}
\end{table}

\subsection{Classification}\label{sec:shclass}
Classification is a mechanism by which group membership labels are assigned to 
unlabelled observations (the group may be a class or cluster). 
Unsupervised classification, or clustering, is the special case 
where all observations are \textit{a~priori} unlabelled or treated as such. 
A common definition of a cluster suggests that it occurs when observations are grouped together in such a way that members of one cluster are more similar to each other than to observations in other clusters. As \cite{mcnicholas16a} mentions, such a definition might be considered flawed because it is satisfied by a solution that places each observation into its own cluster and a definition that casts a cluster as a component in a suitable finite mixture model may be preferable.

Finite mixture models lend themselves well to classification problems. Consider a finite mixture distribution so that $\mathbf{X}$ is a $p$-dimensional random vector that, for all $\vecx \subset \mathbf{X}$, has density of the form
\begin{equation}f(\vecx\mid\varthet)= \sum_{g=1}^G \pi_g f_g(\vecx\mid\vectheta_g),\end{equation}\label{eqn:fmd}where $\pi_g >0$, such that $\sum_{g=1}^G \pi_g = 1$, are the mixing proportions, $f_1(\vecx\mid\vectheta_g),\ldots,f_G(\vecx\mid\vectheta_g)$ are the component densities, and
$\varthet=(\vecpi,\vectheta_1,\ldots,\vectheta_G)$ is the vector of parameters where $\vecpi=(\pi_1,\ldots,\pi_G)$.
Historically, in clustering applications, it has been customary to assume that the component densities are multivariate Gaussian \citep{wolfe65} --- as an aside, it is notable that much recent work focuses on mixtures of non-Gaussian distributions \citep[recent examples include work by][]{bagnato17,wallace18,pesevski18,lin18,morris19,tortora19,wei20}. The Gaussian mixture density can be written 
\begin{equation}\begin{split}\label{eqn:gmm}
f(\vecx\mid\varthet)= \sum_{g=1}^G \pi_g\phi(\vecx\mid\vecmu_g,\matsig_g),
\end{split}\end{equation}
where $\phi(\vecx\mid\vecmu_g,\matsig_g)$ is the multivariate Gaussian density with mean $\vecmu_g$ and covariance matrix~$\matsig_g$. 

Consider a clustering scenario so that we have unlabelled data $\vecx_1,\ldots,\vecx_n$, i.e., there are no known group membership labels. In this case, the Gaussian mixture model likelihood can be expressed as
\begin{equation}
\mathcal{L(\varthet\mid\vecx)}=\prod_{i=1}^n\sum_{g=1}^G
\pi_g\phi(\vecx_i\mid\vecmu_g,\matsig_g),\end{equation}\label{eqn:gmml}where $\pi_g$ can be thought of as the \textit{a priori} probability that observation $\vecx_i$ belongs in component~$g$ \citep{mclachlan00b,mcnicholas16a}.
To enable clustering, the notation $z_{ig}$ is used to represent component (group) membership and has the same definition as before; see \eqref{eqn:thisone1}.
Within the EM algorithm framework, at each iteration, the $z_{ig}$ are replaced by their conditional expected values given the observed data, using the current estimates for the unknown parameters, and we denote this quantity
\begin{equation}\label{zhat}
\hat{z}_{ig}\colonequals\mathbb{E}[Z_{ig}\mid\vecx_i,\hat{\varthet}]=\text{Pr}[Z_{ig}=1\mid\vecx_i,\hat{\varthet}]=\frac{\hat\pi_g\phi(\vecx_i\mid\hat\vecmu_g,\hat\matsig_g)}
{\sum_{h=1}^{G}\hat\pi_h\phi(\vecx_i\mid\hat\vecmu_h,\hat\matsig_h)}.
\end{equation} 

Once parameter estimation has been carried out, the predicted group memberships are obtained from the \textit{a posteriori} probability that observation $\vecx_i$ belongs to component~$g$ --- this is just given by $\hat{z}_{ig}$ evaluated at the (final) parameter estimates, see~\eqref{zhat}. Depending on the problem at hand, this probability may be reported as it is, in which case it is a soft classification, or the probability may be rounded to $0$ or $1$, i.e., hardened. Hardening is usually carried out by computing maximum \textit{a~posteriori} (MAP) classifications, i.e., $\text{MAP}\{\hat{z}_{ig}\}= 1$ if $g=\arg\max_k\{\hat{z}_{ik}\}$ and $\text{MAP}\{\hat{z}_{ig}\}= 0$ otherwise. Note that MAP classifications are reported for the EM algorithms in Section~\ref{sec:eaexp}.
Crucially, regardless of whether $\hat{z}_{ig}$ or $\text{MAP}\{\hat{z}_{ig}\}$ is ultimately returned, the parameter estimation algorithms most often used permit values $\hat{z}_{ig}\in[0,1]$ as the algorithm iterates. This is true for the EM algorithm and some popular alternatives, including variational Bayes approximations \citep[see, e.g.,][]{mcgrory07,subedi14,subedi20}. The EA approach that is taken herein uses values $\tilde{z}_{ig}\in\{0,1\}$ as the algorithm iterates, which is a fundamental difference between it and the more common approaches.

For model-based clustering, the Bayesian information criterion \citep[BIC;][]{schwarz1978} is a commonly used technique for determining the number of components $G$ (if unknown). We can write
\begin{equation}\label{eqn:bic}
\text{BIC} = 2l(\hat\varthet) - \rho\log n,
\end{equation}
where $\hat{\varthet}$ is the maximum likelihood estimate of $\varthet$, $l(\hat\varthet)$ is the maximized log-likelihood, $\rho$ is the number of free parameters in the model, and $n$ is the number of observations. \citet{leroux92}, \cite{roeder97}, \cite{kass95b} and \cite{dasgupta98}, amongst others, provide arguments that support using the BIC for estimating the number of components in a mixture model. 
Alternatives have been suggested for model selection in model-based clustering, \citep[see, e.g.,][]{biernacki00}, but the BIC remains the most common approach. 

\subsection{Model-Based Clustering and $k$-Means Clustering}
As mentioned in Section~\ref{sec:intro}, \cite{celeux92} show that the $k$-means clustering algorithm is equivalent to a CEM algorithm for model-based clustering using a Gaussian mixture model with equal mixing proportions and component covariance structure \begin{equation}\label{eqn:kmeanscov}\matsig_g=\lambda\ident_p,\end{equation} where $\lambda\in\mathbb{R}^+$ and $\ident_p$ is the $p$-dimensional identity matrix. The CEM algorithm uses estimates $\hat{z}_{ig}\in\{0,1\}$ throughout the algorithm and maximizes the classification likelihood \citep[see][Section~2.21 for further details]{mclachlan00b}. Although introduced well over a decade ago, the CEM algorithm never caught on because of the seemingly exaggerated difficulties associated with local maxima. Of particular concern is its propensity for getting ``stuck" at local maxima --- the EM algorithm suffers from the same problem but to a lesser extent. For a summary on the drawbacks of the CEM algorithm, see \citet[][Section~2.21]{mclachlan00b}.

\begin{figure}[!ht]
\centering\includegraphics[width=0.55\textwidth]{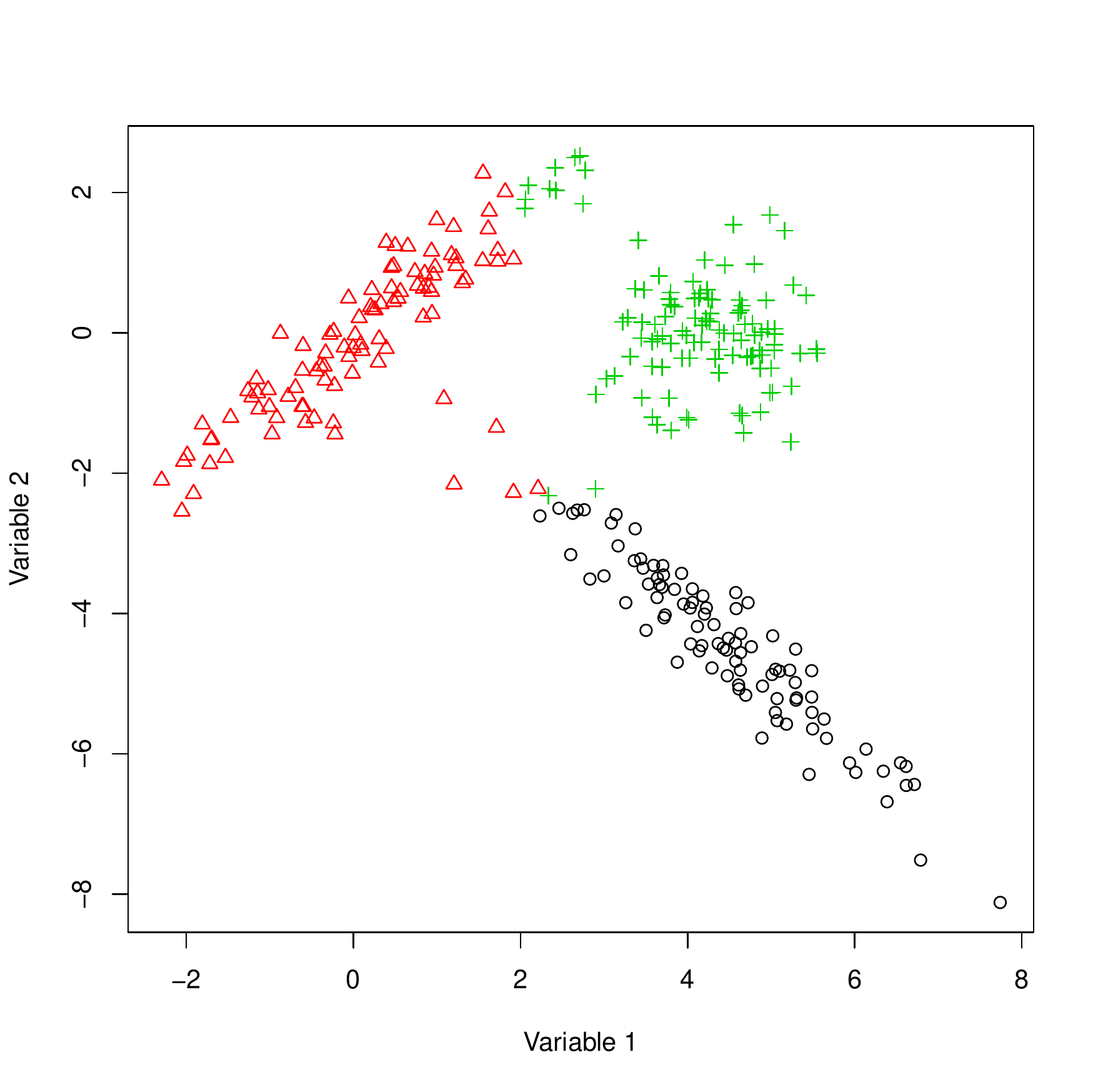}
\vspace{-0.12in}
\caption{Scatterplot of the {\tt x2} data, where plotting symbols and colours reflect the predicted classifications from $k$-means clustering.}\label{fig:x2kmeansfirst}
\end{figure}

Note that the component covariance structure in \eqref{eqn:kmeanscov} limits the associated Gaussian mixture model --- and, by extension, $k$-means clustering --- to spherical components of equal radius. Accordingly, such an approach will only be effective if the clusters are either roughly spheres of equal radius or well separated. Further to the latter situation, clustering where the clusters are well separated can be considered a trivial case and warrants no further consideration. The {\tt x2} dataset from the {\tt mixture} package \citep{browne15b} for {\sf R} \citep{R15} is an example of a seemingly easy clustering situation where $k$-means clustering does not work well. The results for $k$-means are shown in \figurename~\ref{fig:x2kmeansfirst}.

The failure of $k$-means depicted in \figurename~\ref{fig:x2kmeansfirst} is directly attributable to the fact that, unless the clusters are well separated, it accommodates spherical clusters (of equal radius) and is not related to the fact that it is a hard clustering technique. 
Herein, an evolutionary algorithm is used to develop a hard clustering approach, based on a Gaussian mixture model, where clusters have more flexibility in shape, volume, and orientation. Accordingly, in addition to being an alternative approach for parameter estimation in model-based clustering, this work can be viewed as an extension of $k$-means clustering to cases where the components need not be spherical.

\section{Methodology}
\subsection{Model and Fitness Function}
A mixture of multivariate Gaussian distributions is selected as the base model for the EA developed herein. 
As usual, $z_{ig}$ is used to denote component membership labels; see \eqref{eqn:thisone1}. Because we are working within a clustering paradigm, all component memberships are \textit{a~priori} unknown or treated as such. 
While the EM algorithm maximizes the conditional expected value of the complete-data log-likelihood at each iteration, the EA developed herein has a fitness function based on the (observed) log-likelihood
\begin{equation}\label{eqn:cdllB}
l(\varthet)=\sum_{i=1}^{n}\log\left\{\sum_{g=1}^{G}\pi_g\phi(\vecx_i\mid\vecmu_g,\matsig_g)\right\},
\end{equation}
where $\varthet=(\pi_1,\ldots,\pi_G,\vecmu_1,\ldots,\vecmu_G,\matsig_1,\ldots,\matsig_G)$ denotes the model parameters.

As our EA progresses, the estimated value of $z_{ig}$ evolves. To avoid confusion with the expected values $\hat{z}_{ig}$ used in the EM algorithm (see\ Section~\ref{sec:shclass}), we continue to use $\tilde{z}_{ig}$ to denote the estimate of $z_{ig}$ used in our EA. Accordingly, the estimated component membership of $\vecx_i$ in our EA is given by $\tilde{\vecz}_i=(\tilde{z}_{i1},\ldots,\tilde{z}_{iG})$ for $\tilde{z}_{ig}\in\{0,1\}$. The fitness function is just the log-likelihood  \eqref{eqn:cdllB} evaluated at the estimates
\begin{equation}\label{eqn:eaups}
\tilde{\pi}_g=\frac{n_g}{n}, \qquad
\tilde{\vecmu}_g=\frac{1}{n_g}\sum_{i=1}^n\tilde{z}_{ig}\vecx_i, \qquad
\tilde{\matsig}_g=\frac{1}{n_g}\sum_{i=1}^n\tilde{z}_{ig}(\vecx_i-\tilde{\vecmu}_g)(\vecx_i-\tilde{\vecmu}_g)',
\end{equation}
where ${n}_g=\sum_{i=1}^n\tilde{z}_{ig}$.

\subsection{Evolutionary Algorithm}
In our EA, a number of single parents are used and each is cloned many times, with the cloned children reproducing as discussed here. 
For each child, i.e., each clone of a single parent, two observations are chosen at random and their $\tilde{\vecz}_{i}$ values are swapped. There is a check in the code, so that the swap only occurs if the estimated group memberships are different, i.e., if the $\tilde{\vecz}_{i}$ are different; if not, other observations are selected until two different $\tilde{\vecz}_{i}$ are found.
The children of the different single parents never interbreed with each other. After one instance of crossover has been carried out on each cloned child, all of the children (plus the original few single parents) are put into one list in descending order of fitness. The top few are selected to become the new generation of single parents. 
This crossover procedure will help avoid stopping at local maxima of the fitness surface, i.e., the likelihood surface. However, crossover alone will not suffice in clustering applications --- to see why this is so, consider that it is not always possible to improve a clustering result by just swapping the membership label from one observation with that from another --- so a mutation step is also carried out at each iteration. These iterations, of crossover followed by mutation, are repeated until our EA stagnates. 

To crystallize the exact procedure followed in our EA, pseudocode is provided (Algorithm~\ref{algo}). Note that the code used herein was written in {\sf R} and comments within the pseudocode in Algorithm~\ref{algo} use the {\sf R} comment style, i.e., {\tt \#}.
Note also that, after the very first crossover step, the {\tt parents} are simply the best {\tt pars} elements in terms of fitness. Furthermore, note that {\tt compute fitness} entails computing the estimates in \eqref{eqn:eaups} and then computing the log-likelihood \eqref{eqn:cdllB}. The greedy nature of the mutation step is clear; for a given parent, once a mutation increases the fitness, our EA moves on to the next parent. There is a nice general interpretation to this EA: the crossover step provides diversity while the mutation step allows fitness (clustering) improvements that cannot be facilitated by crossover alone. The effectiveness of our EA for traversing the fitness (likelihood) surface is illustrated in Section~\ref{sec:eaexp}.
\begin{algorithm}
\caption{Pseudocode for an EA for model-based clustering.}\label{algo}
\begin{verbatim}
input: x, G, z, pars=2, stagnation, clones
# x is data matrix; G is number of components (groups);  
# z is a list where each element is a matrix of tilde_z_ig values for one parent; 
# pars is the number of parents (defaults to 2); stagnation is the stagnation value; 
# clones is the number of clones
N = number of rows in x
stag=0
while stag < stagnation
    # First, crossover
    for a in 1 to pars
        for b in 1 to clones
            randomly select two unequal labels from parent a
            swap them to get clone (child) b from parent a
            compute fitness for clone (child) b from parent a
        end for
    end for
    sort parents plus children by descending fitness
    # The top four are now the parents
    if top four are unchanged from previous iteration
        stag ++
    else 
        stag = 0
    end if
    # Now, mutation
    for a in 1 to pars
        rand = random permutation of 1,2,...,N
        for i in rand
            swap two distinct elements in label i
            if fitness increases
                break for (i in rand)
            end if
        end for
    end for
    if no mutation has increased log-likelihood
        stag = stag +1
    else 
        sort parents by descending fitness
    end if
end while
return fitness values (log-likelihoods) and labels for the parents
\end{verbatim}\end{algorithm}

\section{Illustrations}\label{sec:eaexp}

\subsection{Overview and Performance Assessment}

The purpose of these illustrations is to compare our EA to two well-established hard clustering approaches, i.e., $k$-means and $k$-medoids, and the EM algorithm for a Gaussian mixture model. For brevity, the latter shall be referred to as ``the EM algorithm" hereafter. To facilitate comparison with our EA, the EM algorithm is run using the {\tt gpcm()} function from the {\tt mixture} package \citep{browne15b} with {\tt mnames="VVV"}. 
First, a simulated dataset is used (Section~\ref{sec:x2}). The remaining analyses (Sections~\ref{sec:voles}--\ref{sec:winedata}) focus on real data. For all analyses, our EAs are run with two parents: one parent is initialized using $k$-means and the other is initialized using $k$-medoids.

Although all of the illustrations in this section are carried out as real cluster analyses (i.e., the data are treated as unlabelled), the true labels are in fact known. Therefore, it is possible to compare the predicted classifications --- for our EA, the final $\tilde{z}_{ig}$ values and, for the EM algorithm, the final $\hat{z}_{ig}$ values --- with the true labels. We carry out this comparison using the adjusted Rand index \citep[ARI;][]{hubert85}, which is the Rand index \citep{rand71} corrected for chance agreement. The Rand index is the ratio of pairwise agreements to total pairs. Arguments as to why the ARI should be used in this circumstance, as opposed to alternatives like the misclassification rate, are given by \cite{steinley04}.

\subsection{{\tt x2} Data}\label{sec:x2}

\begin{figure}[!h]
\vspace{-0.2in}
\centering\includegraphics[width=0.49\textwidth]{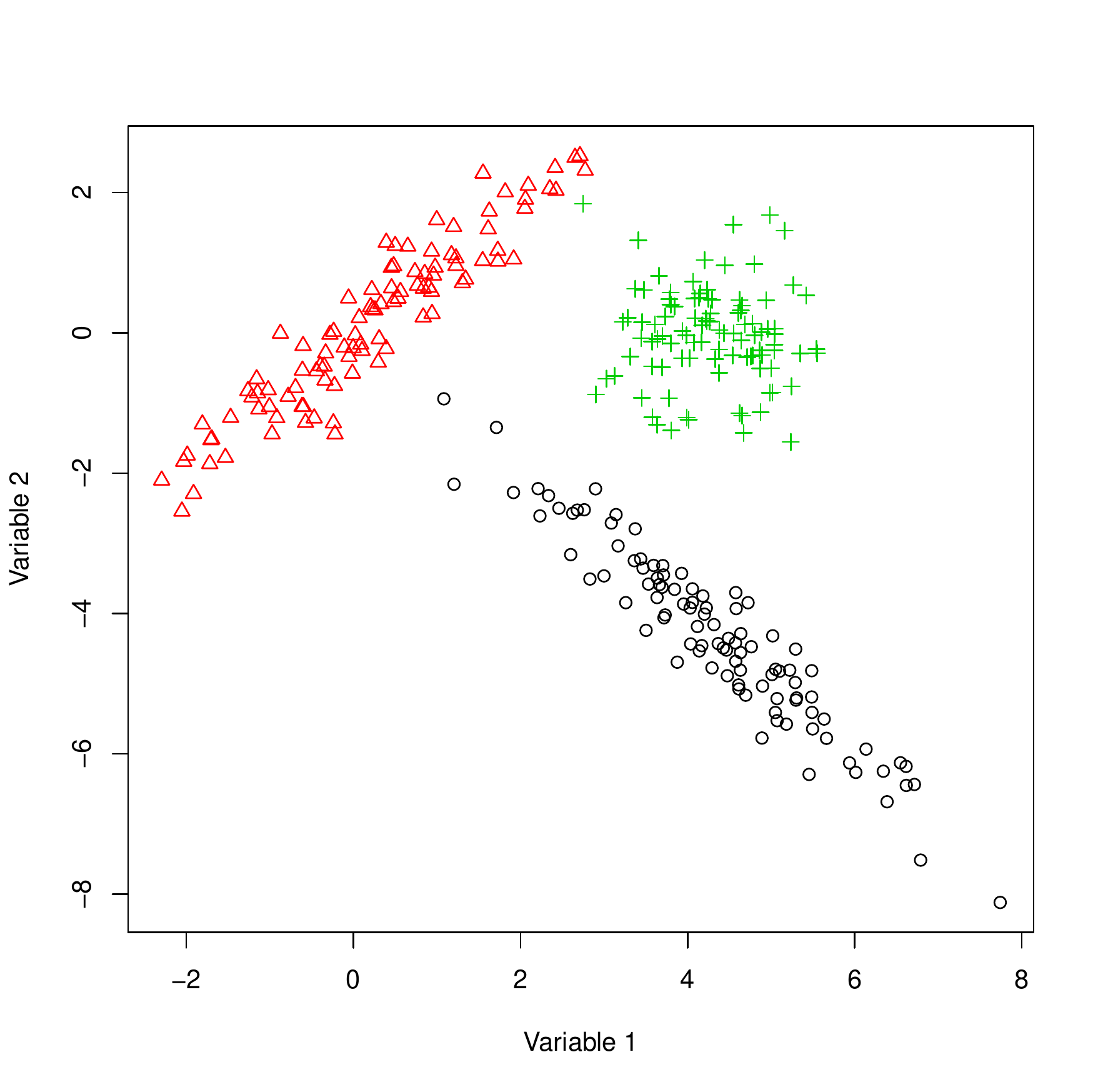} \
\centering\includegraphics[width=0.49\textwidth]{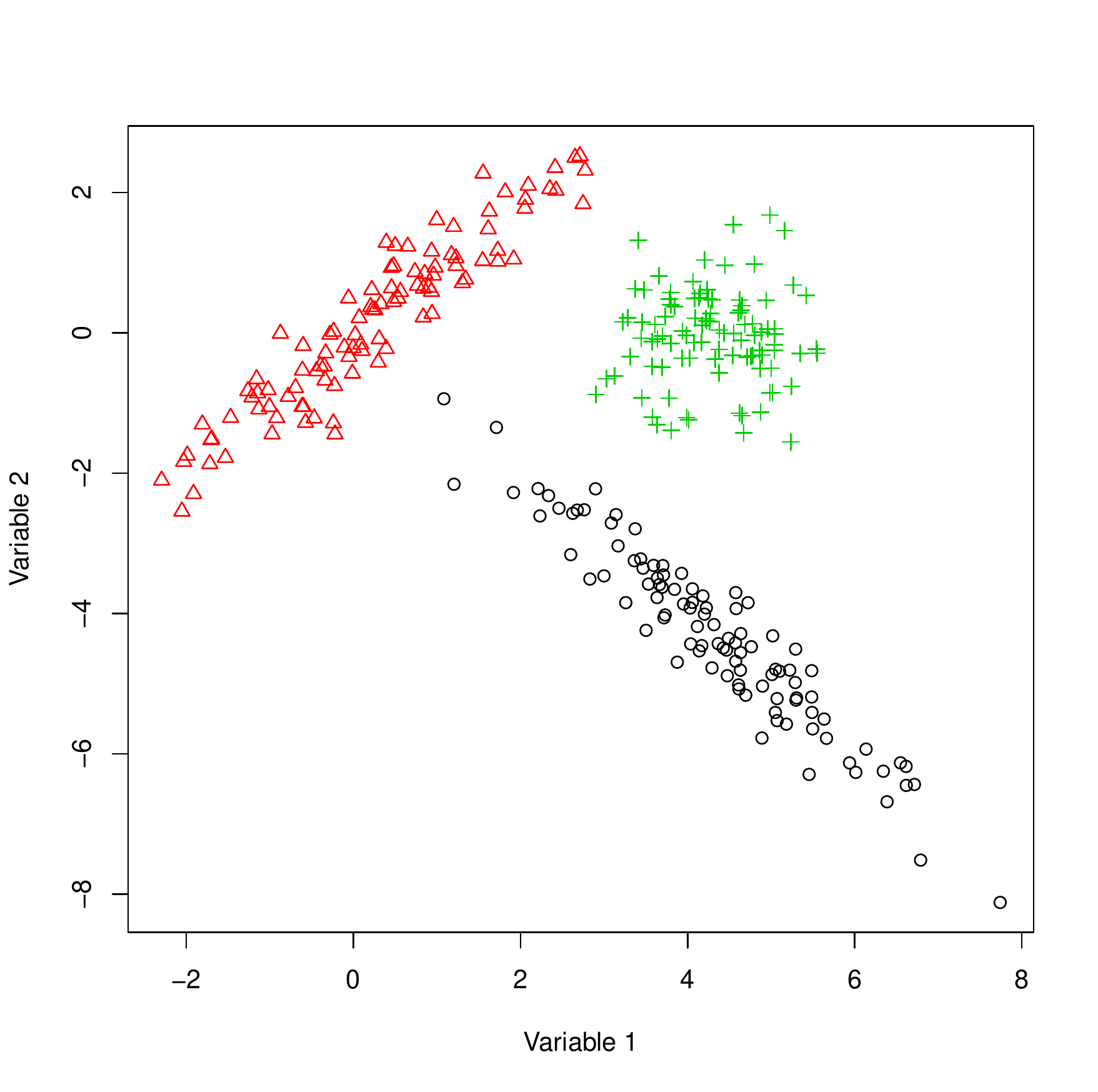}
\vspace{-0.12in}
\caption{Scatterplots of the {\tt x2} data, where plotting symbols and colours reflect the components from the model used to generate the data (left) or the predicted classifications from our EA (right).}\label{fig:x2}
\end{figure}

As a first step, consider the {\tt x2} dataset from the {\tt mixture} package in {\sf R} (\figurename~\ref{fig:x2}). The {\tt x2} data are generated from a Gaussian mixture model with three components and were used by \cite{browne14a} to illustrate their MM (majorization-minimization, in this case) algorithms --- see \cite{hunter04} for an overview of MM algorithms.
As \citet[][Chapter~2]{mcnicholas16a} points out, the {\tt x2} data are a good illustration of data generated from a mixture model where the ``best" clustering result clearly does not correspond perfectly to the labels from the generating model.

\begin{figure}[!h]
\vspace{-0.2in}
\centering\includegraphics[width=0.49\textwidth]{x2kmeans.pdf} \
\centering\includegraphics[width=0.49\textwidth]{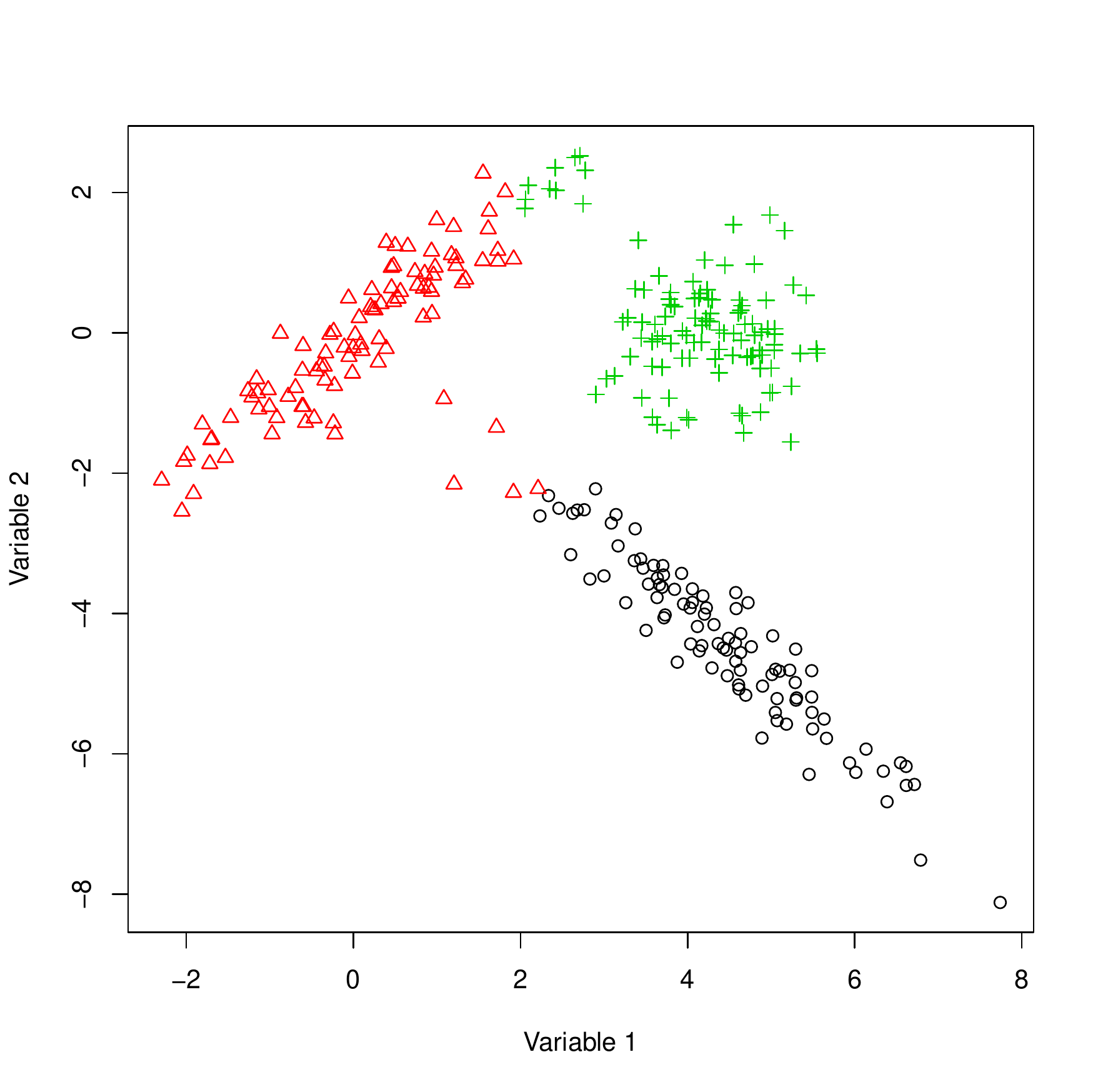}
\vspace{-0.12in}
\caption{Scatterplots of the {\tt x2} data, where plotting symbols and colours reflect the predicted classifications from $k$-means clustering (left) or $k$-medoids clustering (right).}\label{fig:x2kmeans}
\end{figure}

Our EA approach is applied to the {\tt x2} data with the following settings (see Algorithm~\ref{algo}): {\tt stagnation=3} and {\tt clones=10}. The result (\figurename~\ref{fig:x2}) indicates that ``perfect" classification performance is attained --- note, again, that this does not quite correspond to the generating model. It is interesting to consider the performance of other hard clustering approaches and so $k$-means and $k$-medoids are also applied to these data. The results (\figurename~\ref{fig:x2kmeans}) indicate that neither approach performs as well as our EA. Note that the EM algorithm gives the same classification results as our EA for the {\tt x2} data.

\subsection{Female Voles Data}\label{sec:voles}

\begin{figure}[!h]
\vspace{-0.2in}
\centering\includegraphics[width=0.85\textwidth]{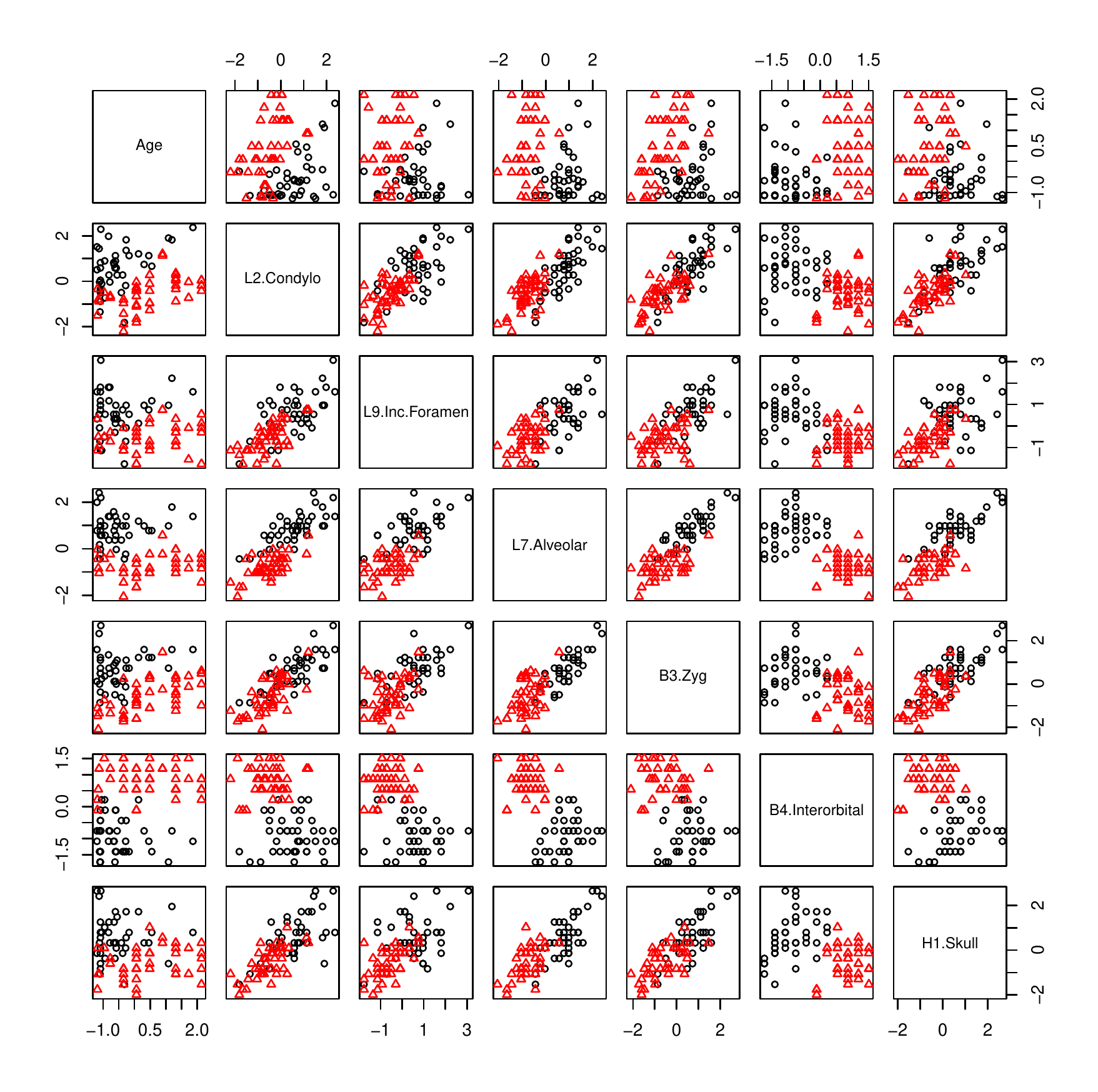}
\vspace{-0.2in}
\caption{Scatterplot of the female voles data, where plotting symbols and colours reflect species.}\label{fig:voles}
\end{figure}

The female voles ({\tt f.voles}) data are available in the {\tt Flury} package \citep{flury12} for {\sf R}. They contain 
six morphometric measurements, as well as age, for 86 female 
voles from two species: \textit{Microtus californicus} and \textit{Microtus ochrogaster} (\figurename~\ref{fig:voles}).

The EA approach introduced herein is applied to these data with {\tt stagnation} $\in\{3,4,5\}$ and {\tt clones} $\in\{10,20,30,40\}$. Over all runs, identical and excellent classification performance was obtained, with just one misclassification (Table~\ref{tab:voles}; $\text{ARI}=0.953$). The respective classification performance of $k$-means (Tables~\ref{tab:voles2}; $\text{ARI}=0.737$) and $k$-medoids (Tables~\ref{tab:voles3}; $\text{ARI}=0.658$) on these data is notably inferior. The EM algorithm gives slightly inferior classification performance, misclassifying one additional vole  
(Table~\ref{tab:voles7}; $\text{ARI}=0.908$) when compared to our EA (Table~\ref{tab:voles}; $\text{ARI}=0.953$).
\begin{table}[!ht]
\caption{Cross-tabulation of the predicted classifications (A, B) from our EA versus species for the  female voles data.}\label{tab:voles}
\vspace{-0.12in}\centering
\begin{tabular*}{0.9\textwidth}{@{\extracolsep{\fill}}lrr}
\hline
 &{A}& {B}\\
\hline
\textit{Microtus californicus} & 41 & 0\\
\textit{Microtus ochrogaster} & 1  & 44\\
\hline
\end{tabular*}
\end{table}
\begin{table}[!ht]
\caption{Cross-tabulation of the predicted classifications (A, B) from $k$-means versus species for the female voles data.}\label{tab:voles2}
\vspace{-0.12in}\centering
\begin{tabular*}{0.9\textwidth}{@{\extracolsep{\fill}}lrr}
\hline
 &{A}& {B}\\
\hline
\textit{Microtus californicus} & 36 & 5\\
\textit{Microtus ochrogaster} & 1  & 44\\
\hline
\end{tabular*}
\end{table}
\begin{table}[!ht]
\caption{Cross-tabulation of the predicted classifications (A, B) from $k$-medoids versus species for the female voles data.}\label{tab:voles3}
\vspace{-0.12in}\centering
\begin{tabular*}{0.9\textwidth}{@{\extracolsep{\fill}}lrr}
\hline
 &{A}& {B}\\
\hline
\textit{Microtus californicus} & 34 & 7\\
\textit{Microtus ochrogaster} & 1  & 44\\
\hline
\end{tabular*}
\end{table}
\begin{table}[!ht]
\caption{Cross-tabulation of the predicted classifications (A, B) from the EM algorithm versus species for the  female voles data.}\label{tab:voles7}
\vspace{-0.12in}\centering
{\small\begin{tabular*}{0.9\textwidth}{@{\extracolsep{\fill}}lrr}
\hline
 &{A}& {B}\\
\hline
\textit{Microtus californicus} & 41 & 0\\
\textit{Microtus ochrogaster} & 2  & 43\\
\hline
\end{tabular*}}
\end{table}

\subsection{Banknote Data}\label{sec:note}

\begin{figure}[!h]
\vspace{-0.2in}
\centering\includegraphics[width=0.65\textwidth]{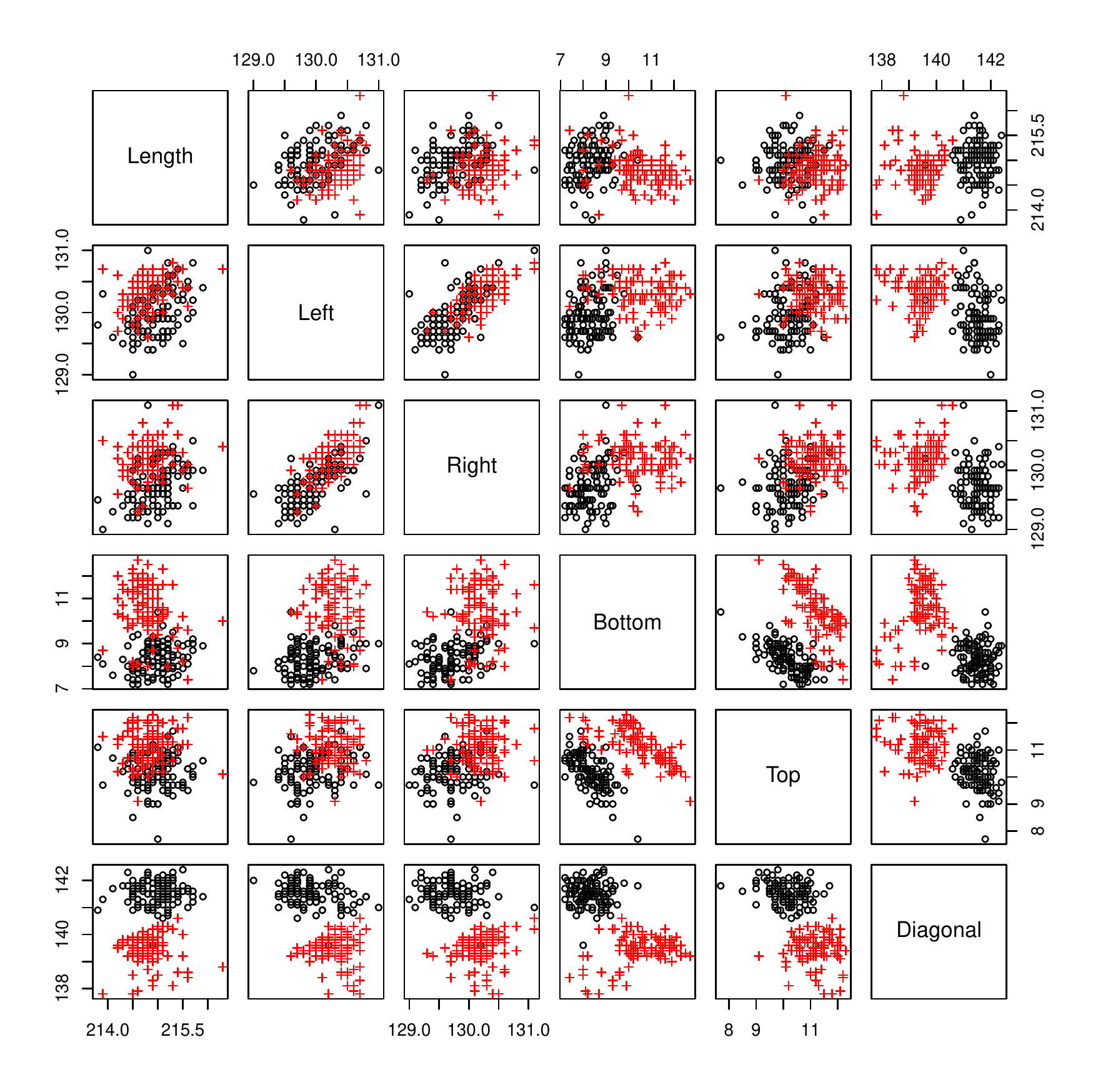}
\vspace{-0.2in}
\caption{Scatterplot of the Swiss banknote data, where plotting symbols and colours reflect true class, i.e., genuine or counterfeit.}\label{fig:notes}
\end{figure}

The {\tt banknote} data are available from the {\tt mclust} package \citep{fraley12b} in {\sf R}. They contain six measurements, all in mm, on 100 genuine and 100 counterfeit Swiss 1000-franc banknotes (\figurename~\ref{fig:notes}).

Our EA is applied to these data with {\tt stagnation} $\in\{3,4,5\}$ and {\tt clones} $\in\{10,20,30,40\}$. Over all runs, identical and excellent classification performance was obtained, with just one misclassification (Table~\ref{tab:notes}; $\text{ARI}=0.980$). The respective classification performance of $k$-means (Table~\ref{tab:notes2}; $\text{ARI}=0.846$) and $k$-medoids (Table~\ref{tab:notes3}; $\text{ARI}=0.941$) on these data is somewhat inferior. The EM algorithm gives the same classification results as our EA on these data (Table~\ref{tab:notes}; $\text{ARI}=0.980$).
\begin{table}[!ht]
\caption{Cross-tabulation of the predicted classifications (A, B) from our EA versus true class for the Swiss banknote data.}\label{tab:notes}
\vspace{-0.12in}\centering
\begin{tabular*}{0.9\textwidth}{@{\extracolsep{\fill}}lrr}
\hline
 &{A}& {B}\\
\hline
Counterfeit & 100 & 0\\
Genuine & 1  & 99\\
\hline
\end{tabular*}
\end{table}
\begin{table}[!ht]
\caption{Cross-tabulation of the predicted classifications (A, B) from $k$-means versus true class for the Swiss banknote data.}\label{tab:notes2}
\vspace{-0.12in}\centering
\begin{tabular*}{0.9\textwidth}{@{\extracolsep{\fill}}lrr}
\hline
 &{A}& {B}\\
\hline
Counterfeit & 100 & 0\\
Genuine & 8  & 92\\
\hline
\end{tabular*}
\end{table}
\begin{table}[!ht]
\caption{Cross-tabulation of the predicted classifications (A, B) from $k$-medoids versus true class for the Swiss banknote data.}\label{tab:notes3}
\vspace{-0.12in}\centering
{\small\begin{tabular*}{0.9\textwidth}{@{\extracolsep{\fill}}lrr}
\hline
 &{A}& {B}\\
\hline
Counterfeit & 100 & 0\\
Genuine & 3 & 97\\
\hline
\end{tabular*}}
\end{table}

\subsection{Italian Wine Data}\label{sec:winedata}

Thus far, the real datasets considered have contained just two classes. To move beyond this, consider the Italian wine data that were collected by \citet{forina86}. A subset of these data, containing  13 chemical and physical properties of three cultivars (Barolo, Grignolino, Barbera) from the Piedmont region of Italy, is available in the {\tt gclus} package \citep{hurley04} for {\sf R}.
\begin{table}[htb]
\caption{The thirteen chemical and physical properties of Italian wines available in {\tt gclus}.}\label{tab:winevars}
\centering
\vspace{-0.12in}\centering
\begin{tabular*}{0.9\textwidth}{@{\extracolsep{\fill}}lll}\hline
Magnesium &  Malic acid & Total phenols\\
Alcohol & Ash &Alcalinity of ash \\
Flavonoids & Nonflavonoid phenols& Proanthocyanins\\
Color Intensity & Hue & Proline\\
OD$_{280}$/OD$_{315}$ of diluted wines & &\\\hline
\end{tabular*}
\end{table}

The EA approach introduced herein is applied to these data with {\tt stagnation} $\in\{3,4,5\}$ and {\tt clones} $\in\{10,20,30,40\}$. Over all runs, identical and excellent classification performance was obtained, with just one misclassification (Table~\ref{tab:wines}; $\text{ARI} = 0.982$). The respective classification performance of $k$-means (Tables~\ref{tab:wines2}; $\text{ARI}=0.897$) and $k$-medoids (Tables~\ref{tab:wines3}; $\text{ARI}=0.741$) on these data is notably inferior.
For the Italian wine data, the EM algorithm gives slightly inferior classification performance (Table~\ref{tab:wines7}; $\text{ARI} = 0.945$) compared to our EA (Table~\ref{tab:wines}; $\text{ARI} = 0.982$).
\begin{table}[!ht]
\caption{Cross-tabulation of the predicted classifications (A, B, C) from our EA versus true class for the Italian wine data.}\label{tab:wines}
\vspace{-0.12in}\centering
{\small\begin{tabular*}{0.9\textwidth}{@{\extracolsep{\fill}}lrrr}
\hline
 &{A}& {B}& {C}\\
\hline
Barolo & 59 & 0 & 0\\
Grignolino & 1  & 70& 0\\
Barbera & 0  & 0& 48\\
\hline
\end{tabular*}}
\end{table}
\begin{table}[!ht]
\caption{Cross-tabulation of the predicted classifications (A, B, C) from $k$-means versus true class for the Italian wine data.}\label{tab:wines2}
\vspace{-0.12in}\centering
{\small\begin{tabular*}{0.9\textwidth}{@{\extracolsep{\fill}}lrrr}
\hline
 &{A}& {B}& {C}\\
\hline
Barolo & 59 & 0 & 0\\
Grignolino & 3  & 65& 3\\
Barbera & 0  & 0& 48\\
\hline
\end{tabular*}}
\end{table}
\begin{table}[!ht]
\caption{Cross-tabulation of the predicted classifications (A, B, C) from $k$-medoids versus true class for the Italian wine data.}\label{tab:wines3}
\vspace{-0.12in}\centering
{\small\begin{tabular*}{0.9\textwidth}{@{\extracolsep{\fill}}lrrr}
\hline
 &{A}& {B}& {C}\\
\hline
Barolo & 59 & 0 & 0\\
Grignolino & 15  & 55& 1\\
Barbera & 0  & 0& 48\\
\hline
\end{tabular*}}
\end{table}
\begin{table}[!ht]
\caption{Cross-tabulation of the predicted classifications (A, B, C) from the EM algorithm versus true class for the Italian wine data.}\label{tab:wines7}
\vspace{-0.12in}\centering
{\small\begin{tabular*}{0.9\textwidth}{@{\extracolsep{\fill}}lrrr}
\hline
 &{A}& {B}& {C}\\
\hline
Barolo & 59 & 0 & 0\\
Grignolino & 3  & 68& 0\\
Barbera & 0  & 0& 48\\
\hline
\end{tabular*}}
\end{table}

\section{Discussion}\label{ch6disc}
An EA has been introduced for model-based clustering. Each iteration of our EA uses a crossover step followed by a (greedy) mutation step; no comparable approach has been taken for Gaussian mixture models. In fact, the closest approach uses mutations only, ignoring crossover \citep[see][]{evol1}.
The clustering philosophy associated with our approach is that of hard clustering, i.e., $\tilde{z}_{ig}\in\{0,1\}$. At no point in our parameter estimation scheme do we entertain soft values. This is in contrast to the commonly used EM algorithm, which uses $\hat{z}_{ig}\in[0,1]$. 

In terms of future work, there is much that could be done. For one, our EA uses an unconstrained (VVV) covariance structure and the other 11 covariance structures used by \cite{celeux95} could easily be implemented --- for practical reasons, comparison with algorithms other than the EM might be preferable for two of these 11 \citep[see][for alternative algorithms]{browne14a,browne14b}. For each of the toy examples (Sections~\ref{sec:voles}--\ref{sec:winedata}), our EA returned the same results for the different values of {\tt stagnation} and {\tt clones} used; however, one would not expect that to be the case in general and so further consideration of model selection is needed. Comparison of our EA to the EM algorithm was limited in scope, i.e., we considered only classification performance, and extent; accordingly, a more in-depth study is required. Extension to higher dimensional problems should be considered and could be achieved in a number of ways, including by extending our EA to mixtures of factor analyzers \citep{ghahramani97,mclachlan00a,mcnicholas08,mcnicholas10d}. It would also be of interest to consider how an EA approach similar to the one introduced herein would work within the mixture discriminant analysis \citep[see][]{scott92,mclachlan92,fraley02a}, semi-supervised \citep[see, e.g.,][]{dean06,mcnicholas10c} or fractionally supervised classification \citep[see][]{vrbik15,gallaugher19a} frameworks. Although our EA is based on a Gaussian mixture model, an analogous approach could be taken for mixtures of non-Gaussian distributions and this will be a topic for future work. Another avenue for future work is using an analogous EA approach within the matrix variate mixture setting, where there has been significant work of late \citep{gallaugher18a,gallaugher19c,gallaugher20,melnykov18,melnykov19,sarkar20,murray20}. Finally, it will be of interest to compare the EA approach developed herein to the variational approximations-deviance information criterion rubric used by \cite{subedi20}.

\section*{Acknlwledgements}
This work was supported by an Alexander Graham Bell Scholarship from the Natural Sciences and Engineering Research Council of Canada (NSERC; S.M.~McNicholas). This work was partly supported by the Canada Research Chairs program (P.D.~McNicholas) and respective NSERC Discovery Grants (S.M.~McNicholas, P.D.~McNicholas, D.A.~Ashlock).


\end{document}